# Temperature dependence of resistivity and Hall-coefficient in a strongly disordered metal: NbN

Madhavi Chand[a], Archana Mishra[a,b], Y. M. Xiong[c], Anand Kamlapure[a], S. P. Chockalingam[a], John Jesudasan[a], Vivas Bagwe[a], Mintu Mondal[a], P. W. Adams[c], Vikram Tripathi[a] and Pratap Raychaudhuri[a1]

[a]*Tata Institute of Fundamental Research, Homi Bhabha Rd., Colaba, Mumbai 400005, India.*
[b]*IIC, Indian Institute of Technology Roorkee, Roorkee, Uttarakhand 247667, India.*
[c]*Department of Physics and Astronomy, Louisiana State University, Baton Rouge, Louisiana 70803, USA*

*Abstract:* We report the temperature dependence of resistivity ($\rho$) and Hall coefficient ($R_H$) in the normal state of homogeneously disordered epitaxial NbN thin films with $k_F l$~3.27-10.12. The superconducting transition temperature ($T_c$) of these films varies from 8.13K to 16.8K. While our least disordered film displays usual metallic behavior, for all the films with $k_F l \leq 8.13$, both $\frac{d\rho}{dT}$ and $\frac{dR_H}{dT}$ are negative up to 300K. We observe that $R_H(T)$ varies linearly with $\rho(T)$ for all the films with $\left(\frac{R_H(T)-R_H(300K)}{R_H(300K)}\right)=\gamma\left(\frac{\rho(T)-\rho(300K)}{\rho(300K)}\right)$, where $\gamma$=0.68±0.1. Measurements performed on a 2nm thick Be film shows similar behavior with $\gamma$=0.69. This behavior is inconsistent with existing theories of localization and *e-e* interactions in a disordered metal.

[1] Electronic mail: pratap@tifr.res.in

## I. Introduction

The evolution of electronic properties with disorder remains one of the most challenging problems of modern condensed matter physics[1]. In recent years, experimental investigations in this field have received renewed interest motivated by several novel phenomena observed in strongly disordered superconductors, such as magnetic field driven superconductor insulator transitions in disordered $InO_x$ and TiN films[2,3] close to the superconductor-insulator transition, unusual temperature dependence of normal state resistivity[4] in underdoped High $T_c$ cuprates, and the formation of a pseudogapped normal state[5,6] in strongly disordered conventional superconductors. For weak disorder in a normal metal, scattering from disorder potential results in an increase in the temperature independent part of the electrical resistivity, whereas the temperature dependent part remains largely unaltered. However, it was shown by Anderson[7] that in the presence of strong disorder, all the electronic states in a 3 dimensional (3D) metal can get localized giving rise to an insulating state. However, experimentally determining the exact disorder level where the Anderson metal-insulator transition happens in a 3D disordered system is often complicated due the presence of strong electron-electron (*e-e*) interactions close to the metal insulator transition. *e-e* interactions caused by the diffusive motion of the electrons can give rise to non-trivial temperature dependent corrections to the resistivity (ρ) even change the sign[8,9] of $\frac{d\rho}{dT}$ while the system is a metal. It is therefore important to investigate the evolution of the transport properties as a function of disorder strength to understand the role of *e-e* interactions in the presence of strong disorder.

Theoretically, in addition to $\rho$, the Hall coefficient ($R_H$) is another quantity that is expected to get modified due to *e-e* interactions[10]. In the weak scattering regime the relative correction to $\rho$ and $R_H$ is predicted to be related by[11], $\frac{\Delta R_H}{R_H} = 2\frac{\Delta\rho}{\rho}$. In this paper we investigate the temperature dependence of $\rho$ and $R_H$ in 3-dimensional disordered epitaxial NbN thin films with[12] $k_F l$~3.27-10.12. All these films have a sharp superconducting transition ($\Delta T_c$<0.3K) at low temperatures with the superconducting transition temperature ranging between 8.13K and 16.8K. The thickness of these films ($t$ > 50nm) is much larger than $\xi_0$~5nm and the electronic mean free path ($l$~2-4Å). With increasing disorder the transport properties of these films evolve from conventional metallic behavior (e.g. $\frac{d\rho}{dT}$ > 0, $R_H$ independent of $T$) for our most ordered films to an unusual metallic state where $\frac{d\rho}{dT}, \frac{dR_H}{dT} < 0$. The main result of this paper is that for all the films with $\frac{d\rho}{dT}, \frac{dR_H}{dT} < 0$, $R_H(T)$ varies linearly with ρ(T) over a large range of temperature. In addition, we observe a universal relation, $\left(\frac{\Delta R_H}{R_H(285K)}\right) = \gamma\left(\frac{\Delta\rho}{\rho(285K)}\right)$ ($\Delta R_H = R_H(T) - R_H(285\text{K})$ and $\Delta\rho = \rho(T) - \rho(285\text{K})$) with $\gamma\approx 0.685$, in sharp contrast to the prediction in the weak scattering regime. For comparison, similar measurements done on ultrathin Be films show a very similar behavior with $\gamma\approx 0.69$.

## II. Experimental Details

Epitaxial NbN films (similar to the ones used in ref. 13,) were grown through reactive magnetron sputtering on (100) oriented single crystalline MgO substrate, by sputtering a Nb target in $Ar/N_2$ gas mixture. The effective disorder was controlled by changing the sputtering power or the $Ar/N_2$ ratio which effectively changed the $Nb/N_2$ ratio in the plasma. X-ray diffraction and transmission electron microscope studies reveal that all our films are epitaxial with no extrinsic source of physical granularity. The 2nm thick Be film was deposited on polished glass substrate using e-beam deposition in an initial vacuum of ~0.1μTorr. The Be film also showed no salient morphological features when investigated with scanning force microscope. Further details of sample preparation and characterization have been reported in ref., 14 and 15. Resistance, magnetoresistance and Hall measurements were performed in a home built cryostat up to a maximum field of 12T. $\rho$ and $R_H$ at different temperatures were measured using a standard four-probe ac technique on films patterned in a Hall bar geometry using a shadow mask. $R_H$ was calculated from Hall voltage deduced from reversed field sweeps from +12 to −12 T after subtracting the resistive contribution. The thickness of the films was determined within an accuracy of 15% using a stylus profilometer by measuring on different parts of the patterned sample.

**III. Results**

Figure 1 shows the cross sectional TEM images of two films with the highest and lowest level of disorder. The epitaxial nature of the films is evident from the NbN lattice planes which match with the lattice planes of the MgO substrate. This is also seen from X-ray diffraction studies[16]. This implies that the increase in disorder does not destroy the epitaxial nature of our films. Therefore disorder in our films does not result from physical granularity but rather from atomic scale disorder such as vacancies in the crystalline lattice. Figure 2(a) shows $\rho(T)$ as a function of temperature for films with different $k_F l$ up to 300K. An expanded view of $\rho(T)$-$T$ for the film with the highest $T_c$ (showing conventional metallic behavior), is shown in the inset. All the films with $k_F l \leq 8.13$ show a negative $d\rho/dT$ extending up to room temperature. The temperature dependence of $R_H(T)$ in the same temperature range extracted from the slope of $\rho_{xy}$- $B$ curves (Figures 2(c)-(d)), is shown in Figure 2(b). The central result of this paper, namely, the linear variation of $R_H$ as a function of $\rho$ obtained using the temperature variation of $\rho(T)$ and $R_H(T)$, is shown in figures 3(a)-(c). This linear variation is also evident from the inset of Figure 3(a) where temperature variation of $\rho(T)$ and $R_H(T)$ are shown in the same graph. Within error bars of our measurement, $R_H$ follows a relation $R_H(T)=R_{H0}+A\rho(T)$ for all the disordered films.

We now turn our attention to the precise temperature dependence of $\rho(T)$ for the disordered samples. For the more disordered samples, $\sigma(T)$ varies linearly with $T$ from 40K to 150K (Figure 4(a)) and as $T^{1/2}$ (Figure 4(b)) at higher temperatures. It is interesting to note that the slopes of the linear parts of the $\sigma$-$T$ and $\sigma$-$T^{1/2}$ curves are independent of the $k_Fl$ within error bars. All our films show an upward deviation from the linear $T$ behavior at temperatures below ~40K. While the onset of superconductivity and the very high upper critical field precludes the possibility of investigating the normal state conductivity down to very low temperatures, the trend in variation of $\sigma(T)$-$T$ suggests that $\sigma(T)\neq 0$ for $T\rightarrow 0$. Similar behavior in the resistivity has earlier been reported[17] in 3-D disordered $In_2O_{3-x}$ films. Since our measurements are done up to moderately high temperatures it is important to assess the role of electron-phonon (*e*-ph) scattering in $\sigma(T)$ at high temperatures. To estimate this contribution in the inset of Figure 4(a) we show an expanded view of $\rho(T)$-$T$ of the sample with $k_Fl$~8.82. The $\rho(T)$-$T$ curve has a shallow minimum around 240K where $d\rho/dT$ changes from negative to positive. The positive slope at high temperature which can be attributed to the *e*-ph scattering, results in a resistivity increase of 0.01μΩ m in the interval 240K and 300K. Therefore the contribution of *e*-ph scattering to the overall $\rho(T)$ is less than 2% for this sample. For more disordered samples this contribution is therefore negligible compared to impurity scattering. It is also worthwhile to explore whether the upward deviation of $\sigma(T)$ at low temperatures from the linear $T$ behavior in the highly disordered films is due to proximity to the superconducting transition. Since superconducting fluctuations[18] can be suppressed by the application of magnetic field we measured the magnetoresistance (MR) at temperatures above $T_c$ (Figure 4(c) inset). While the downturn in $\rho(T)$ just before $T_c$ is suppressed by the application of magnetic field (Figure 3(c)), for T>30K the MR is negligible. It is therefore likely that the deviation from $\sigma(T) \propto T$ behavior for T<40K is an intrinsic property of the normal state and not associated with onset of superconductivity.

## IV. Discussion

We now compare our data with existing predictions for disordered metals. A power law dependence of $\sigma(T)$ on $T$ in a disordered metal has been predicted to happen both from weak localization as well as *e-e* interactions. However, for a 3-D disordered system without *e-e* interactions, scaling theory[19] predicts that $R_H$ will retain its metallic character down to the metal-insulator transition and will be temperature independent. *e-e* interactions on the other hand can change both the value of $\rho$ and $R_H$. In the weak scattering regime in 3-D, *e-e* correlation predicts a conductivity of the form' ,

$$\sigma = \sigma_0 + \frac{e^2}{\hbar}\frac{1}{4\pi^2}\frac{1.3}{\sqrt{2}}\left(\frac{4}{3} - \frac{3}{2}\tilde{F}_\sigma\right)\sqrt{\frac{T}{D}} \quad , (1)$$

where $D(=\frac{v_F^2\tau}{d}$ where d is the dimensionality) is the diffusivity and $\tilde{F}_\sigma$ is a Fermi liquid parameter such that $\left(\frac{4}{3} - \frac{3}{2}\tilde{F}_\sigma\right)$ is of the order of unity. This is similar to the observed temperature dependence of our disordered films over a large range of temperature. There are, however, two difficulties in reconciling with this scenario. First, calculation of the prefactor of $T^{1/2}$ in equation (1) using the experimental value[20] of $v_F$ (=1.758×10$^6$m s$^{-1}$) and $\tau$ (=1.227×10$^{-16}$s) for the most disordered sample shows that it is two orders of magnitude smaller than the corresponding experimental values of the slope obtained from the slope of $\sigma(T)$ vs. $T^{1/2}$ curve in the temperature range where it is linear. More importantly, it has been shown that for *e-e* interactions the relative correction in $R_H$ is double[21] that of $\rho$, namely,

$$\frac{\Delta R_H}{R_H} = 2\frac{\Delta \rho}{\rho}. \quad (2)$$

In Figure 5(a) we plot $\frac{\Delta R_H}{R_H}\left(=\frac{R_H(T)-R_H(285K)}{R_H(285K)}\right)$ vs. $\frac{\Delta\rho}{\rho}\left(=\frac{\rho(T)-\rho(285K)}{\rho(285K)}\right)$ for all NbN samples with $k_Fl$≤8.13. All the curves are linear within error bars of our measurements and the slope, $\gamma = d\left(\frac{\Delta R_H}{R_H}\right)\Big/ d\left(\frac{\Delta\rho}{\rho}\right)$ has a universal value 0.68±0.1. This is in clear contradiction with eqn. 2. While equation (2) is strictly valid for the limit H→0, taking this limit would not alter our results since for our samples $\rho_{xy}$ is linear over the entire magnetic field range (Figures 1(c)-(d)). The inset of Figure 5(a) shows that within error bars, γ is independent of $k_Fl$.

We now consider possible scenarios where γ can deviate from 2. Since for localization effects γ=0 and for *e-e* interaction γ=2, γ can in-principle take any intermediate value in the presence of both localization and *e-e* interactions. This scenario, however, can be ruled out for two reasons. First, since for samples with larger $k_Fl$, *e-e* interactions will be more predominant than localization effects, a systematic deviation of γ towards 2 should have been observed with increasing $k_Fl$. Secondly, in 3-D the temperature dependence of σ(T) due to localization and *e-e* interactions have different temperature dependence. Since $R_H$ is affected only by *e-e* interactions, the linear relation between σ(T) and $R_H$ is not expected over a large temperature range. The second possibility is that our films are not in the weak scattering regime ($k_F l$>>1) for which equations 1 and 2 are applicable. However, since the level of disorder in the films shown in Fig.5 spans a large range of $k_F l$~3.27-8.13, one would have again expected a systematic variation of $\gamma$ towards the theoretically expected value of 2 for the samples with larger $k_F l$. Such a systematic change was actually observed in a 2-D electron gas[22] in Si inversion layer where a gradual increase of $\gamma$ towards the theoretical value of 2 was observed as the sheet resistance $R_{sq}\rightarrow 0$. No such systematic variation is observed in our data. To check if the observed behavior is specific to NbN, similar measurements were also performed on a 2nm thick Be film (Figure 5(b)) with sheet resistance, $R_{sq}\approx 3.61$ kΩ. While this film displays a logarithmic temperature dependence of ρ(T) typical of a disordered 2-D system, the slope $\gamma = d\left(\frac{\Delta R_H}{R_H}\right) \Big/ d\left(\frac{\Delta R_{sq}}{R_{sq}}\right) \approx 0.69$, is strikingly similar to the value observed in disordered NbN. It would be interesting to investigate if the value of $\gamma$ in the 2D Be films is robust as a function as a function disorder ($R_{sq}$) similar to NbN.

In this context we would like to note that the theoretical value of $\gamma\approx 2$ has so far been observed in 2D electron gas in Si inversion layers[23] in the limit of large sheet resistance and at intermediate magnetic fields (~0.1-0.5T), whereas $\gamma$ decreases from 2 for both very low fields as well as higher field values. The former is attributed to the dominance of localization at very low fields whereas the latter arises presumably due to breakdown of the low field limit where equation 2 is valid. The first effect is also observed in 3-D disordered $In_2O_3$ films[24] where $R_H$ measured at very low magnetic field was reported to be temperature independent. While the very low limit of $R_H$ is below our experimental resolution we do not observe any non-linearity in $\rho_{xy}$ vs. H at fields above ~0.2T. On the other hand, Hall effect measurements on uncompensated Si:As samples[25] in the metallic regime showed that $\gamma$ ranges 0.4–0.6 for different samples. We therefore believe that further theoretical considerations may be needed to establish the value of $\gamma$ in a disordered metallic system.

Since our measurements can be done only down to the superconducting transition temperature of the NbN thin films, we could also consider the possibility that our more disordered samples are actually in the insulating regime, where the temperature dependence of $\sigma(T)$ and $R_H(T)$ are expected to show similar behavior[26]. In that case it is expected that $\sigma(T)$ will go to zero at a temperature below our minimum temperature of measurement. While such a possibility cannot be ruled out it is clear from the data that this would require a sharp downturn in the conductivity for T<12K even for the most disordered sample. This would signify the onset of a new energy scale which is smaller than our lowest temperature of measurement (~1meV). Therefore, even if such an energy scale exists it cannot manifest itself over the large temperature range of our observations.

**V. Conclusions**

In summary, we report the unusual temperature dependence of $\rho$ and $R_H$ in the normal state of homogenously disordered epitaxial NbN thin films with the disorder level varying from moderately clean to very dirty limit. All the samples are in the metallic regime as seen by a finite $\sigma(T)$ as $T \rightarrow 0$. The films with $k_F l$<8.13 display a large negative $d\rho/dT$ extending up to room temperature. A quantitative analysis shows that the overall change in conductivity with temperature is much larger than what is expected from weak localization or *e-e* interactions. We find a remarkably linear relation between $R_H(T)$ and $\rho(T)$ extending from low temperature up to 285K in all the films with $\frac{d\rho}{dT} < 0$. While a linear relation between $\rho(T)$ and $R_H(T)$ is in agreement with the corrections in $\rho$ and $R_H$ arising from *e-e* interactions, the relative change in the two quantities follow the relation $\frac{\Delta R_H}{R_H(285K)} = (0.68 \pm 0.1)\frac{\Delta\rho}{\rho(285K)}$ independent of $k_F l$. Measurements on ultrathin Be films suggest that this behavior is not specific of NbN but is a more general property of disordered metals. We believe that our results represent a fundamental contradiction with the weak scattering prediction of *e-e* interactions in the disordered metallic systems.

***Acknowledgements:*** We would like to thank T. V. Ramakrishnan, V. F. Gantmakher and Z. Ovadyahu for useful discussions, D. Shahar for critical remarks and D. Khmel'nitzkii for his clarification on the validity of equation (2) in 3-D. We would also like to thank B. A. Chalke, R. D. Bapat and S. C. Purandare for technical help with the TEM. Y.M. Xiong and P.W. Adams acknowledge the support of the U.S. DOE under Grant #DE-FG02-07ER46420.

## Appendix I

In our earlier publications (ref ,) dealing with the superconducting state of disordered NbN, we have reported the $k_F l$ measured from the value of $\rho$ and $R_H$ at 18K. In this paper we report the $k_F l$ values determined from the corresponding quantities at 285K. This is because at higher temperature there are no corrections in $\rho$ and $R_H$ due to localization or e-e interaction[27] up to the leading order in $1/k_F l$. In Table 1 we provide a look-up table that gives the $k_F l$ values calculated at 18K and 285 K for the same samples in order to have a correspondence with references , .

**Table 1: Look-up table for $k_F l$ values calculated at 18K and 285K.**

| $k_F l$ (18K) | $k_F l$ (285K) |
|---|---|
| 1.54 | 3.27 |
| 2.14 | 3.65 |
| 3.33 | 4.98 |
| 4.08 | 5.5 |
| 6.36 | 8.01 |
| 7.47 | 8.13 |
| 8.38 | 8.82 |
| 12.07 | 10.12 |

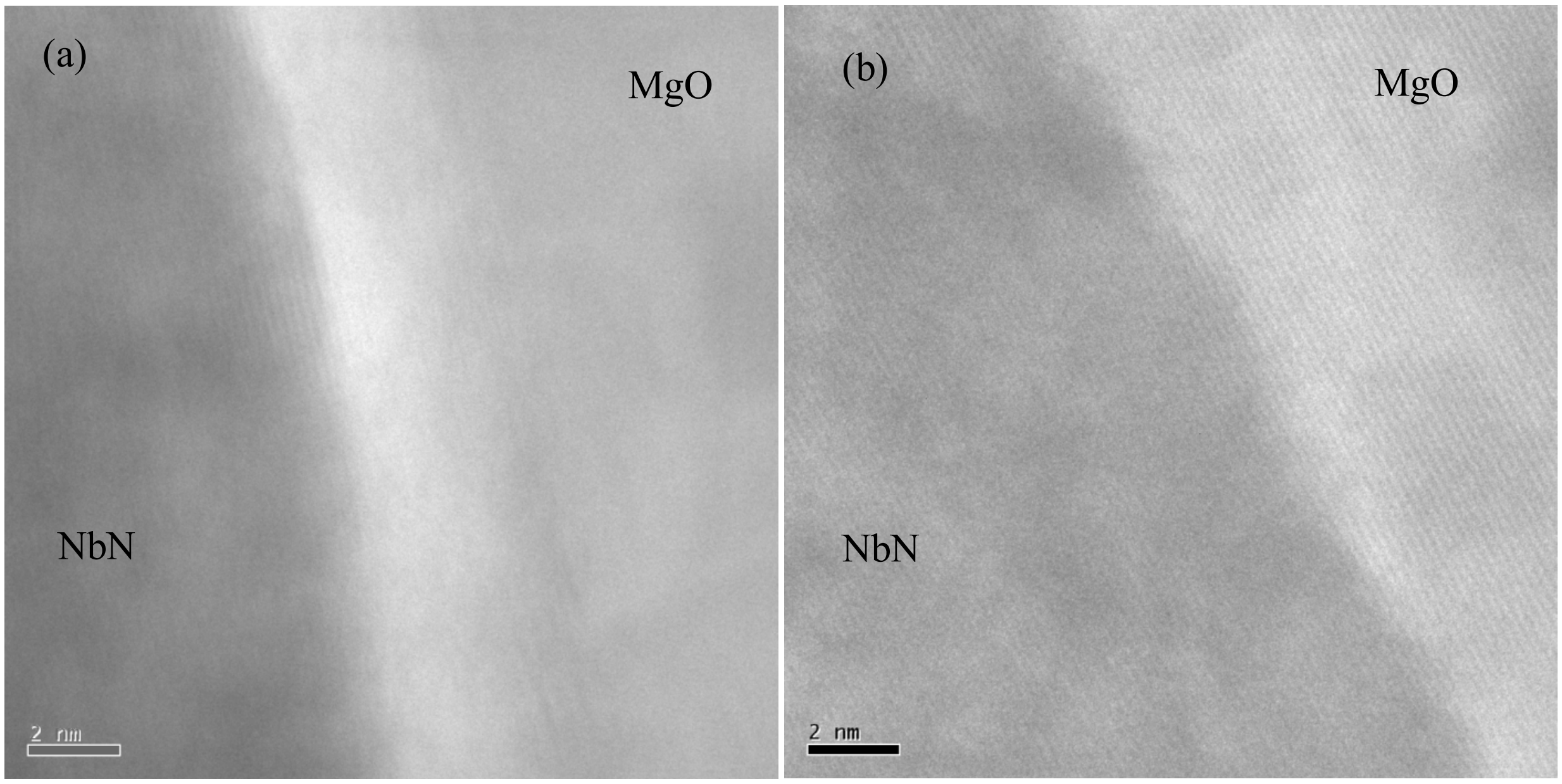

**Figure 1 (a) Transmission Electron Micrograph of disordered NbN film with a $k_F l$~3.3; (b) Transmission Electron Micrograph of an NbN film with $k_F l$~9.**

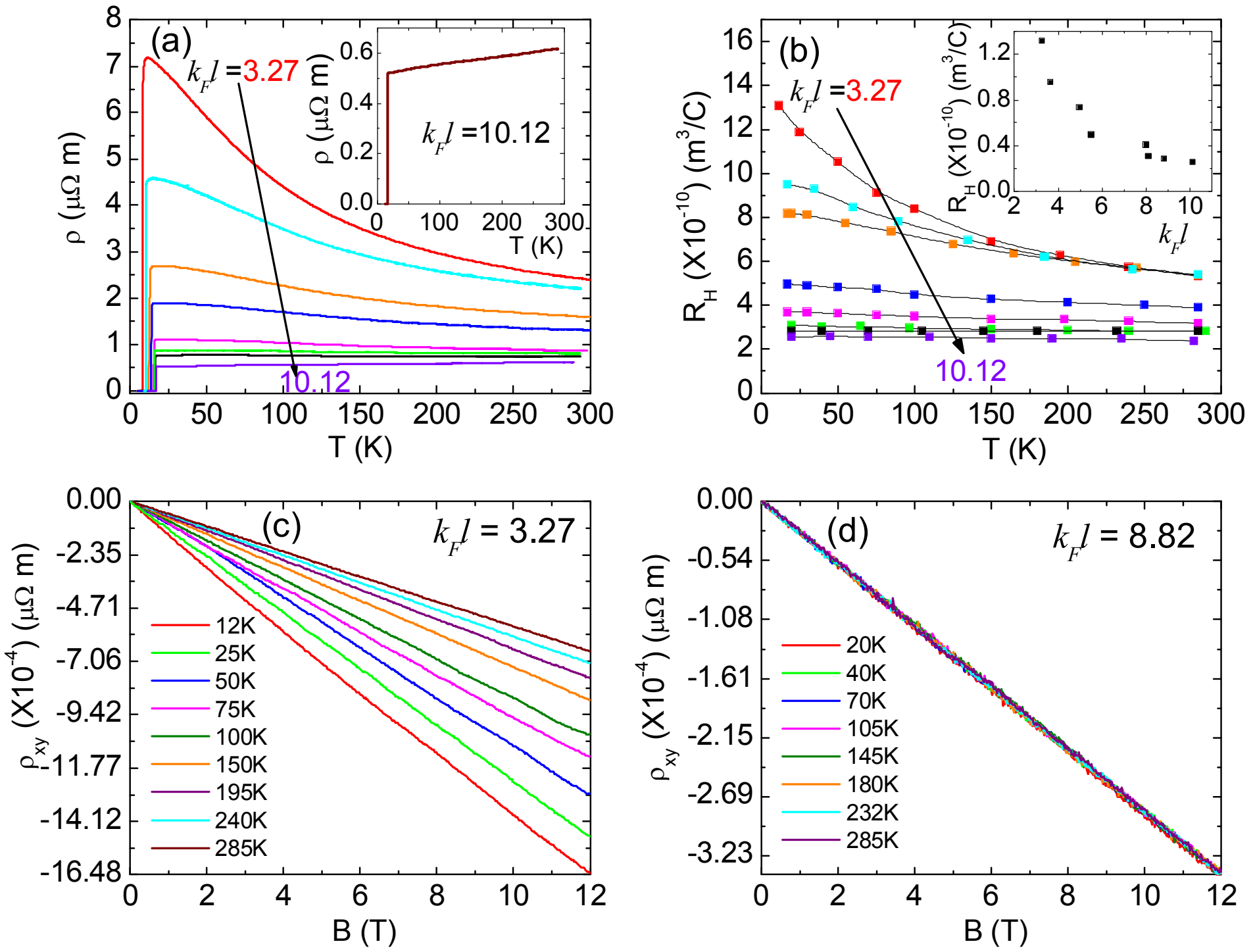

**Figure 2: (a) $\rho(T)$ vs. T for the films of different $k_Fl$ and (inset) an expanded view of the $\rho(T)$ for the most ordered sample with $k_Fl$~10.12; (b) $R_H$ vs. $T$ for all the films and (inset) $R_H$ as a function of $k_Fl$ at 18K;The arrows are in the direction of increasing $k_Fl$ values: 3.27, 3.65, 4.98, 5.50, 8.01, 8.13, 8.82 and 10.12; (c) and (d): Hall resistivity vs. magnetic field at different temperatures for the samples with $k_Fl$ ~3.27 and 8.82 respectively.**

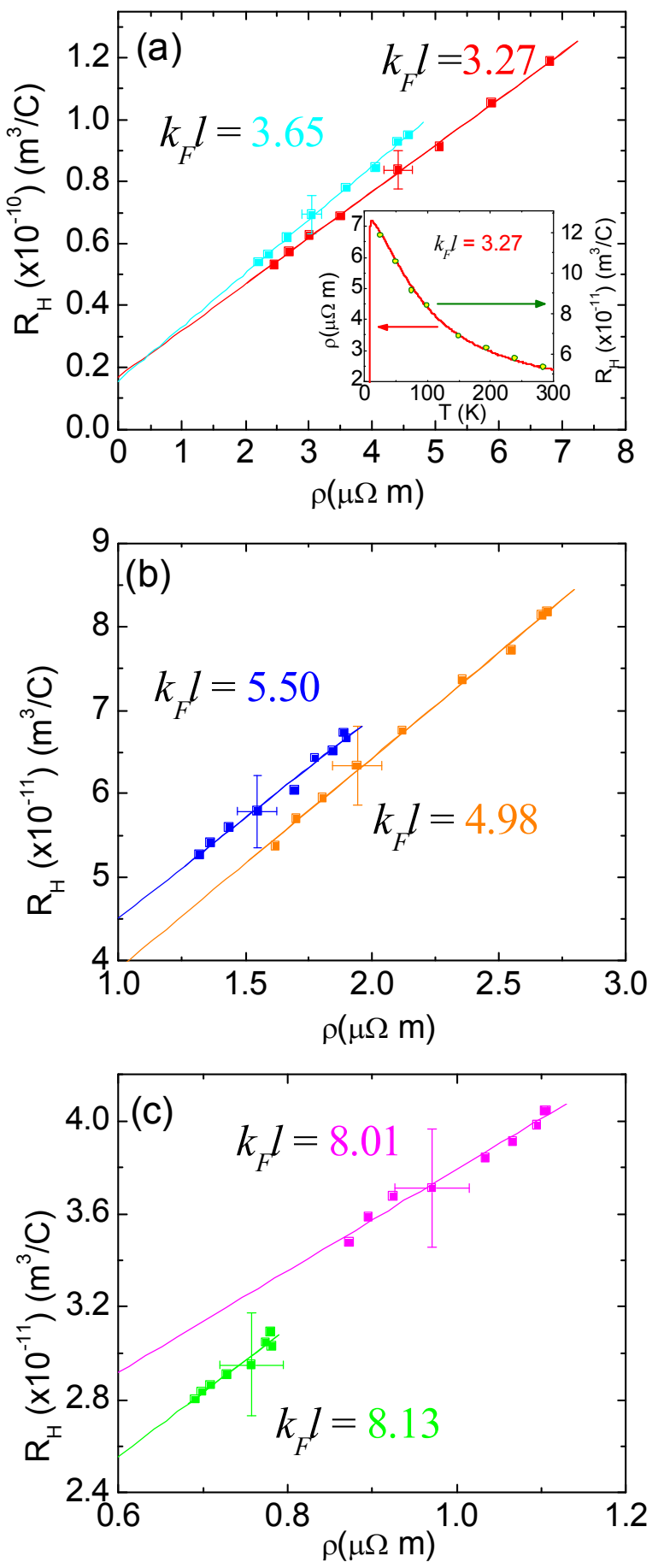

**Figure 3: $R_H(T)$ vs. $\rho(T)$ for the samples with (a) $k_Fl$ ~ 3.27 and 3.65, (b) $k_Fl$ ~ 4.98 and 5.50 and (c) $k_Fl$ ~ 8.01 and 8.13; The inset in (a) shows $R_H(T)$ superposed on the corresponding $\rho(T)$ displaying the linear relationship**

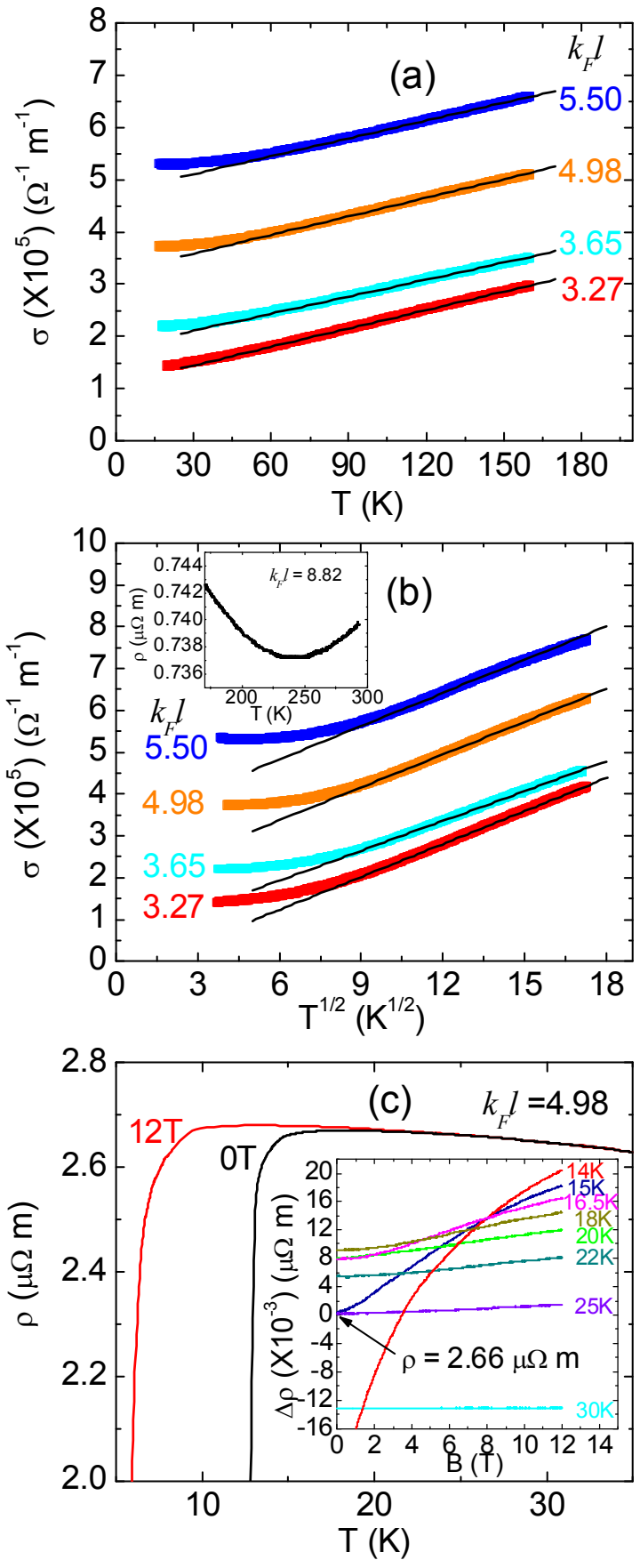

**Figure 4 : (a) $\sigma$ vs. $T$ for the disordered films showing the linear behavior at lower temperatures. The average slope of the linear fits is 1145.34 $\Omega^{-1}$ $m^{-1}$/K; (b) $\sigma$ vs. $T^{1/2}$ along with linear fits with the mean slope 25606.92 $\Omega^{-1}$ $m^{-1}/K^{1/2}$. In both (a) and (b) all the slopes are within 8% of the mean value. The inset of (b) shows an expanded $\rho(T)$ for the film with $k_Fl$ ~ 8.82; (c) $\rho(T)$ in zero field and in 12T for the film with $k_Fl$ ~ 4.98 and (inset) the MR for the same film at different temperatures.**

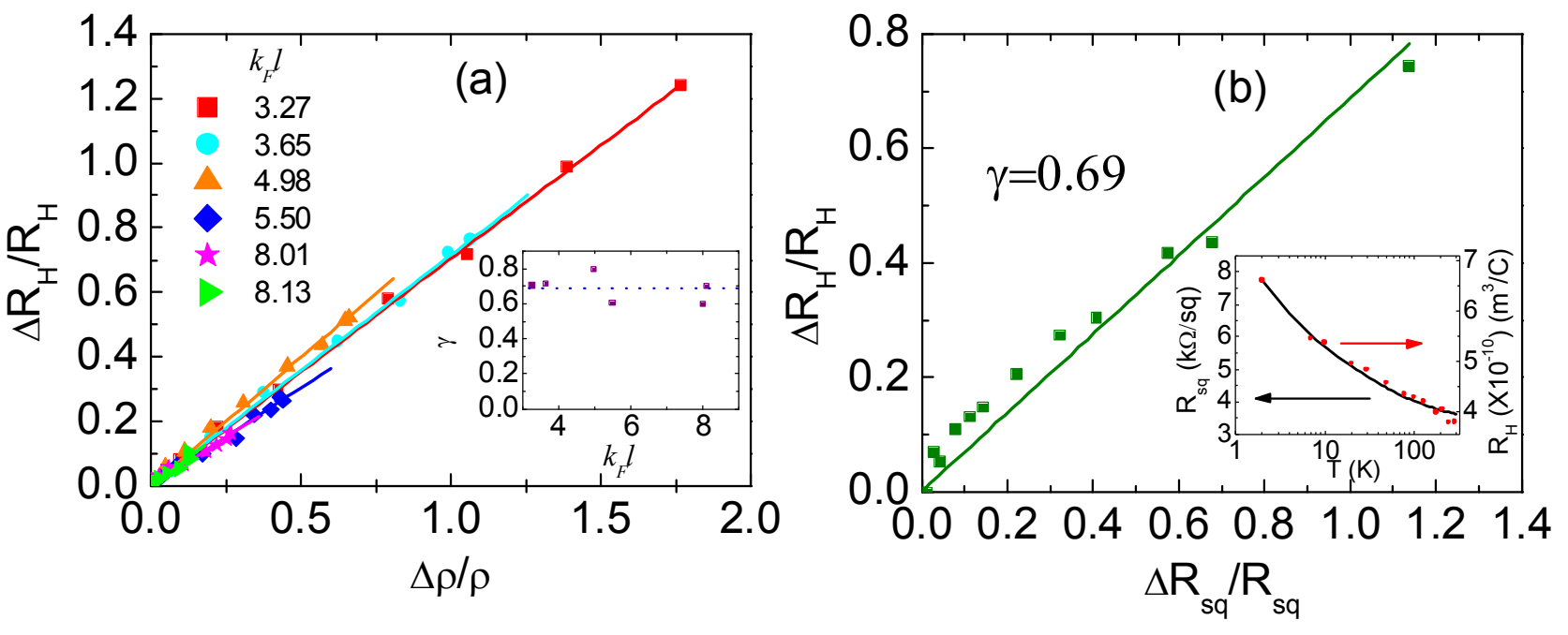

**Figure 5: (a)** $\frac{\Delta R_H}{R_H}\left(=\frac{R_H(T)-R_H(285K)}{R_H(285K)}\right)$ $vs.\frac{\Delta\rho}{\rho}\left(=\frac{\rho(T)-\rho(285K)}{\rho(285K)}\right)$ **(shown as points) for the different NbN samples. The linear fits (shown as lines) have an average value of 0.685±0.108; the inset shows** $\gamma\ (= d(\frac{\Delta R_H}{R_H})/d(\frac{\Delta\rho}{\rho}))$ **as a function of** $k_Fl$ **with the average value shown as a dashed line. (b)** $\frac{\Delta R_H}{R_H}\left(=\frac{R_H(T)-R_H(290K)}{R_H(290K)}\right)$ $vs.\frac{\Delta R_{sq}}{R_{sq}}\left(=\frac{R_{sq}(T)-R_{sq}(290K)}{R_{sq}(290K)}\right)$ **for the ultrathin Be film. The γ value is 0.69. The inset shows the temperature variation** $R_H(T)$ **and** $R_{sq}(T)$ **for the ultrathin Be film.**

**References:**

[1] Allen M. Goldman and Nina Marković, Phys. Today 51, 39 (1998).

[2] G. Sambandamurthy, L. W. Engel, A. Johansson, E. Peled, and D. Shahar, Phys. Rev. Lett. **94**, 017003 (2005).

[3] T. I. Baturina, A. Yu. Mironov, V. M. Vinokur, M. R. Baklanov, and C. Strunk, Phys. Rev. Lett. **99** 257003 (2007).

[4] Yoichi Ando , G. S. Boebinger, A. Passner, Phys. Rev. Lett. **75**, 4662 (1995).

[5] B. Sacépé, C. Chapelier, T. I. Baturina, V. M. Vinokur, M. R. Baklanov, and M. Sanquer, Phys. Rev. Lett. **101** 157006 (2008).

[6] S. P. Chockalingam, M. Chand, A. Kamlapure, J. Jesudasan, A. Mishra, V. Tripathi, and P. Raychaudhuri, Phys. Rev. B **79**, 094509 (2009).

[7] P. W. Anderson, Phys. Rev. **109,** 1492 (1958).

[8] B. L. Altshuler, A. G. Aronov and P. A. Lee, Phys. Rev. Lett. **44**, 1288 (1980); B. L. Altshuler and A. G. Aronov, Ch. 1 in *Electron-Electron Interactions in Disordered Systems,* edited by M. Pollak and A. L. Efros (North-Holland, Amsterdam) (1984).

[9] M. Lee, J. G. Massey, V. L. Nguyen and B. I. Shklovskii, Phys. Rev. B **60**, 1582 (1999)

[10] P.A. Lee and T.V.Ramakrishnan, Rev. Mod. Phys. **57**, 287 (1985).

[11] B.L.Altshuler, D. Khmel'nitzkii, A.I. Larkin and P.A.Lee, Phys. Rev. B, **22**, 5142 (1980).

[12] The value of $k_Fl$ is estimated using the formula values $k_Fl=((3\pi^2)^{2/3}\hbar(R_H(285\text{K}))^{1/3})/(\rho(285\text{K})e^{5/3})$. In reference 6 and 13, $k_Fl$ was calculated using the corresponding values of $R_H$ and $\rho$ at 17K. One to one correspondence between these two values is given in Appendix 1.

[13] S. P. Chockalingam, M. Chand, J. Jesudasan, V. Tripathi, and P. Raychaudhuri, Phys. Rev. B **77**, 214503 (2008).

[14] Y. M. Xiong, A. B. Karki, D. P. Young and P. W. Adams, Phys. Rev. B **79,** 020510 (2009).

[15] V.Yu. Butko, J. F. DiTusa, and P.W. Adams, Phys. Rev. Lett. **84,** 1543 (2000).

[16] S P Chockalingam, M. Chand, J. Jesudasan, V. Tripathi and P. Raychaudhuri, J. Phys.: Conf. Ser. **150**, 052035 (2009).

[17] Z. Ovadyahu, J. Phys. C: Solid State Phys. **19**, 5187 (1986).

[18] L.G. Aslamazov and A.I. Larkin, Sov. Phys.-Solid State, 10, 875 (1968), C. Caroli, K. Maki, Phys. Rev. **159**, 306 (1967), R.S. Thompson, Phys. Rev. B, **1**, 327 (1970)

[19] B. Shapiro and E. Abraham, Phys. Rev. B **24**, 4025 (1981).

[20] $v_F$ and $\tau$ are calculated using the formulae $v_F = \hbar k_F / m = \frac{\hbar (3\pi^2 n)^{1/3}}{m}$ and $\tau = m\sigma(285K)/ne^2$ where $n = 1/eR_H(285K)$ and $m$ is the free electron mass.

[21] While in ref. equation (2) has been derived for a 2D system, this relation also remains valid in 3D; D. Khmel'nitzkii (private communication).

[22] D. J. Bishop, D. C. Tsui and R. C. Dynes, Phys. Rev. Lett. **46,** 360 (1981); M. J. Uren, R. A. Davies and M. Pepper, J. Phys. C: Solid St. Phys. **13,** L985 (1980).

[23] R. C. Dynes, Surface Science **113,** 510 (1982).

[24] E. Tousson and Z. Ovadyahu, Solid State Commun. **60,** 407 (1986). $R_H$ was not measured at large fields in this paper.

[25] D. W. Koon and T. G. Castner, Phys. Rev B **41,** 12054 (1990).

[26] L. Friedman, J. Non-Cryst. Solids, 6, 329 (1971)

[27] M. Khodas and A.M. Finkel'stein, Phys. Rev. B **68,** 155114 (2003)